\documentclass[prl,twocolumn,superscriptaddress,showpacs]{revtex4-1}
\usepackage{amsmath}%
\usepackage{amsfonts}%
\usepackage{amssymb}%
\usepackage{graphicx}
\usepackage{color}
\usepackage{tabularx}
\usepackage{braket}

\usepackage[normalem]{ulem} 
\usepackage{color} 

\begin{document}

\title{3D Cavity quantum electrodynamics with a rare-earth spin ensemble}

\author{S.~Probst}
\affiliation{Physikalisches Institut, Karlsruhe Institute of Technology, D-76128 Karlsruhe, Germany}

\author{A.~Tkal\v{c}ec}
\affiliation{Physikalisches Institut, Karlsruhe Institute of Technology, D-76128 Karlsruhe, Germany}

\author{H.~Rotzinger}
\affiliation{Physikalisches Institut, Karlsruhe Institute of Technology, D-76128 Karlsruhe, Germany}

\author{D.~Rieger}
\affiliation{Physikalisches Institut, Karlsruhe Institute of Technology, D-76128 Karlsruhe, Germany}

\author{J-M.~Le Floch}
\affiliation{ARC Centre of Excellence for Engineered Quantum Systems,
University of Western Australia, 35 Stirling Highway, Crawley WA 6009, Australia}

%

\author{M.~Goryachev}
\affiliation{ARC Centre of Excellence for Engineered Quantum Systems, University of Western Australia, 35 Stirling Highway, Crawley WA 6009, Australia}

\author{M.~E.~Tobar}
\affiliation{ARC Centre of Excellence for Engineered Quantum Systems, University of Western Australia, 35 Stirling Highway, Crawley WA 6009, Australia}

\author{A.~V.~Ustinov}
\affiliation{Physikalisches Institut, Karlsruhe Institute of Technology, D-76128 Karlsruhe, Germany}

\author{P.~A.~Bushev}
\affiliation{Experimentalphysik, Universit\"{a}t des Saarlandes, D-66123 Saarbr\"{u}cken, Germany}

\date{\today}

\begin{abstract}
We present cavity QED experiments with an Er$^{3+}$:Y$_2$SiO$_5$~(Er:YSO) crystal magnetically coupled to a 3D cylindrical sapphire loaded copper resonator. Such waveguide cavities are promising for the realization of a superconducting quantum processor. Here, we demonstrate the coherent integration of a rare-earth spin ensemble with the 3D architecture. The collective coupling strength of the Er$^{3+}$ spins to the 3D cavity is 21 MHz. The cylindrical sapphire loaded resonator allowed us to explore the anisotropic collective coupling between the rare-earth doped crystal and the cavity. This work shows the potential of spin doped solids in 3D quantum circuits for application as microwave quantum memories as well as for prospective microwave to optical interfaces.
\end{abstract}

\pacs{42.50.Fx, 76.30.Kg, 03.67.Hk, 03.67.Lx, 76.30.-v}


\maketitle


Today, the field of quantum information science is looking for the possible physical and technological realization of future quantum processors. A considerable attention is focused on the study of isolated quantum systems such as trapped ions, electronic and nuclear spins, optical photons and superconducting~(SC) quantum circuits. A promising route towards the realization of a feasible technology lies in the coherent integration of different systems resulting in \emph{hybrid quantum system}~\cite{Nori2013}. Such a hybrid system will benefit from the best physical features of its isolated parts, as for instance, scalability and rapid manipulation of SC qubits, and long coherence time of atoms~\cite{Tian2004, Verdu2009}.

One way of implementing a hybrid quantum system, is to couple atomic ensembles magnetically to a planar superconducting quantum circuit~\cite{Verdu2009}. Here, the strong confinement of a resonator mode along a coplanar microwave line mediates a strong collective coupling between the spins of the trapped atoms and the SC resonator. In spite of the persisting development of experiments on coupling trapped rubidium atoms to planar SC circuits, the SC hybrid circuits based on trapped atoms are still challenging to realize in practice~\cite{Fortagh2011, Fortagh2013}. In that respect, crystals doped with magnetic ions (nitrogen vacancy centers in diamond or rare-earth ion doped solids) are an appealing alternative atomic system \cite{Imamoglu2009, Shuster2010, Bertet2011, Amsuss2011, DiCarlo2013, Saito2013, Probst2013}. Such solid states spin systems can easily be integrated with various planar SC quantum circuits.

In contrast to the long coherence times of spin systems~\cite{Bertaina2007, Tyryshkin2011}, the coherence of modern SC planar circuits is still limited by few microseconds, due to uncontrollable coupling to the environment~\cite{Houck2008}. The drastic improvement in coherence is possible by introducing a new architecture for SC quantum circuits based on three-dimensional resonators, which has been recently proposed and successfully implemented~\cite{Paik2011, Rigetti2012}. Two-qubit gate operations have been demonstrated~\cite{Poletto2012}, and multi-qubit entanglement schemes in 3D circuit QED have been proposed~\cite{Girvin2013}.

From the perspective of electron spin resonance~(ESR) spectroscopy, 3D cavities are used since the beginning of the field. It is also known, that a paramagnetic material with a very narrow inhomogeneous spin linewidth $\Gamma_2^{\star}/2\pi$ ($\sim100$~kHz at the microwave X-band, ranging from 8-12 GHz) often yields a complex response, see Ref.~\cite{BrookerBook} chapter 6. Such an effect can be explained by the strong coupling of a spin system to a 3D cavity. In fact, strong coupling of spins to a 3D cavity has been recently demonstrated in a conventional room temperature electron spin resonance experiments~\cite{Chiorescu2010, Morton3D}.


In general, the coupling strength of spins to a cavity of any geometry is $v=\textrm{g}\mu_B \sqrt{\mu_0 \omega_0 n_s \xi/4 \hbar}$, where $\textrm{g}$ is electronic \textrm{g}-factor, $\mu_B$ is the Bohr magneton, $\omega_0$ is the resonance frequency, $n_s$ is the spin concentration and $\xi$ is the filling factor~\cite{Bushev2011}. Since the filling factor of a 3D cavity can be quite large (e.g. 20\%, see Ref.~\cite{Tobar2005}), the collective coupling strength $v/2\pi$ can attain $\sim$10 MHz for the usual experimental concentration of spins in solids $n\simeq 10^{17}$cm$^{-3}$. This collective coupling strength of the spins is practically the same as in the case of 2D circuit QED and meets the requirement for coherent strong coupling $v>\kappa,\Gamma^\star_2$, i.e, the coupling has to be larger than the resonator decay rate and the inhomogeneous broadening of the spin ensemble. Therefore, one can implement a quantum memory based on spin doped solids in a three dimensional circuit QED.

In this article, we present a 3D cavity QED experiment with an Er$^{3+}$:Y$_2$SiO$_5$~(Er:YSO) crystal and show collective strong coupling of erbium electronic spins to a copper waveguide resonator at millikelvin temperatures. This crystal is a very promising material for an optical~\cite{Gisin2010} as well as a microwave quantum memory~\cite{Afzelius2013}, because its optical magnetic dipole transition possesses the longest measured optical coherence time of 6 ms at the telecom C-band around 1.54 $\mu m$~\cite{Sun2006}. The detailed exploration of such material for the application in hybrid systems has recently started~\cite{Bushev2011, Wilson2012}, and strong coupling of electronic spins has also been demonstrated~\cite{Probst2013}. Yet, such a crystal is considered as an excellent candidate for reversible coherent conversion of microwave photons into the optical telecom C-band~\cite{Morigi2014, Longdell2014}.
\begin{figure}[htb]
    	\includegraphics[width=0.98\columnwidth]{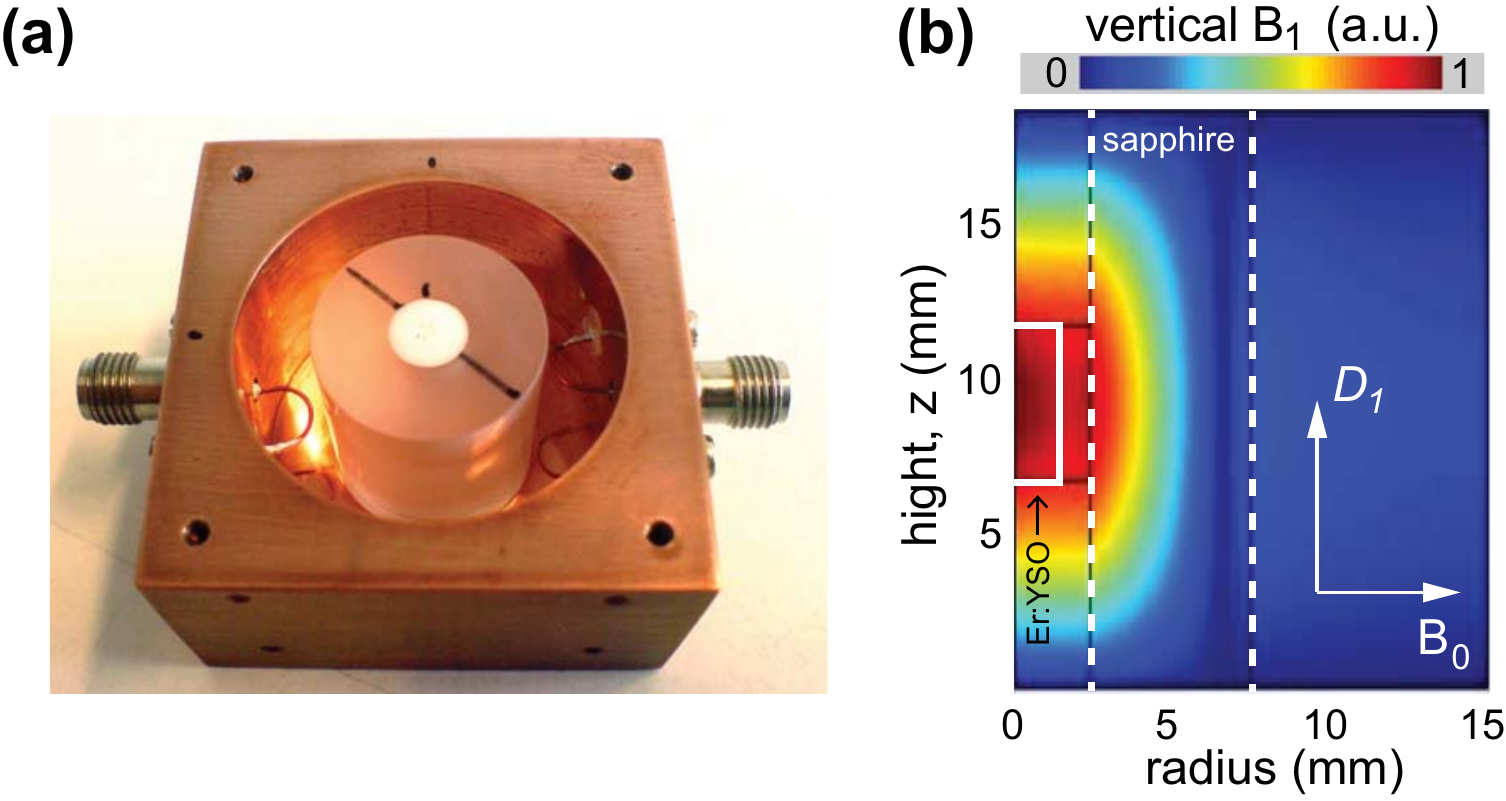}
    	\caption{(Color online)~\textbf{(a)}~Picture of the 3D hybrid quantum system. The Er:YSO crystal of 3 $\times$ 3.5 $\times$ 5 mm$^3$ size is placed inside the sapphire loaded copper resonator. The external magnetic field is applied in the direction perpendicular to the symmetry axis of the cylindrical cavity. \textbf{(b)}~Dimensions of the experiment, orientation of the crystal and simulation of the oscillating magnetic field component $B_1$ of the TE$_{011}$ mode along the symmetry axis of the cavity. }
	\label{fig_setup}
\end{figure}

The picture of the 3D hybrid quantum system is presented in Fig.~\ref{fig_setup}(a). We use a single Y$_{2}$SiO$_{5}$ crystal doped with 0.005\% of Er$^{3+}$ (Er:YSO), supplied by Scientific Materials Inc. The crystal has dimensions of 3~$\times$~3.5 $\times$~5 mm$^3$ and its optical extinction axis $D_1$ is oriented along the 5 mm long side. The crystal is placed inside a sapphire loaded microwave resonator. The resonator is a polished copper cavity with a radius of 15 mm and a hight of 19 mm. The sapphire cylinder has internal and external radii of 6 mm and 9 mm respectively and it is placed inside the copper cavity. The sapphire loaded microwave cavity is operated at the TE$_{011}$ mode at a resonance frequency of $\omega_c/2\pi=5.592$~GHz. Its loaded quality factor can be adjusted by coupling loops in the range between 700 and 70000 at low temperature. Compared to room temperature, the intrinsic quality factor of the resonator at cryogenic temperature shows a four fold increase. At the presented experiment the cavity was overcoupled with the loaded $Q_l=800$, corresponding to a cavity HWHM linewidth of $\kappa_c/2\pi=3.5$~MHz.

Figure~\ref{fig_setup}(b) displays the dimensions of the experiment as well as a simulation of the microwave oscillating field $B_1\sin\omega t$ component, which is directed along the symmetry axis of the cylindrical cavity. The Er:YSO crystal is placed in the center of the sapphire cylinder, where the oscillating field is maximal, and it is kept in the position by two teflon corks. The sapphire cylinder confines the electric field due to its large dielectric constant, which reduces the effective mode volume by factor of approximately 3. The magnetic filling factor of the mode inside the Er:YSO crystal is about $\xi\simeq0.2$. The crystal inside the cavity can be rotated in the $b-D_2$ plane in order to study the anisotropic coupling of Er:YSO. The experiment is placed inside a BlueFors BF-LD-250 dilution refrigerator, and microwave spectroscopy is performed at a base temperature of $T$~=~20~mK.
\begin{figure}[htb]
    	\includegraphics[width=0.98\columnwidth]{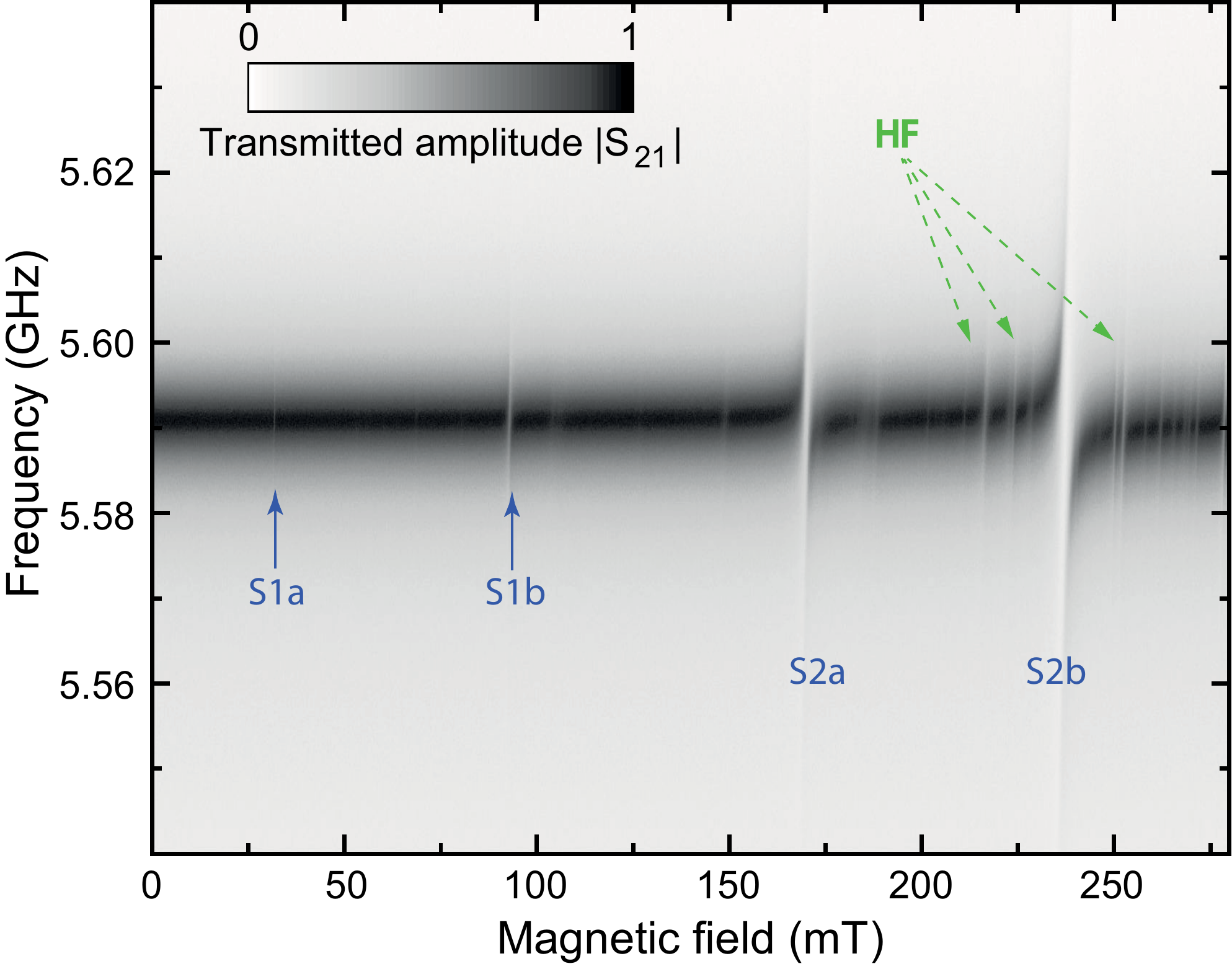}
    	\caption{(Color online)~ESR spectrum of the Er:YSO crystal coupled to the waveguide resonator. The four electronic spin transitions are marked according to their site and class, see the text for explanation and ref.~\cite{Probst2013}. The hyperfine transitions of $^{167}$Er are clearly visible at high fields.}
	\label{fig_spectrum_main}
\end{figure}

The microwave transmission ESR spectrum of the Er:YSO crystal for the applied field region between 0 and 278 mT is shown in Fig.~\ref{fig_spectrum_main}. The probing power of the microwave signal at the entrance of the cavity was set to 100 fW, which corresponds to about 100 microwave photons inside the resonator. The spectrum consist of 4 avoided level crossings due to the coupling of 4 electronic spin transitions. The labels $S_1$ and $S_2$ denote magnetic transitions of Erbium ions occupying site 1 and 2, respectively, and $a, b$ denote the inequivalent magnetic positions (magnetic classes) of the ions in the crystal~\cite{Guillot2006, Sun2008}. Due to the strong magnetic anisotropy, the coupling strength of the transitions increases with the magnitude of the magnetic field $B_0$~\cite{Probst2013}. Therefore, the hyperfine transitions (HF) associated with the presence the $^{167}$Er isotope appear only at high fields in the measured spectrum.

\begin{figure}[htb]
    	\includegraphics[width=0.9\columnwidth]{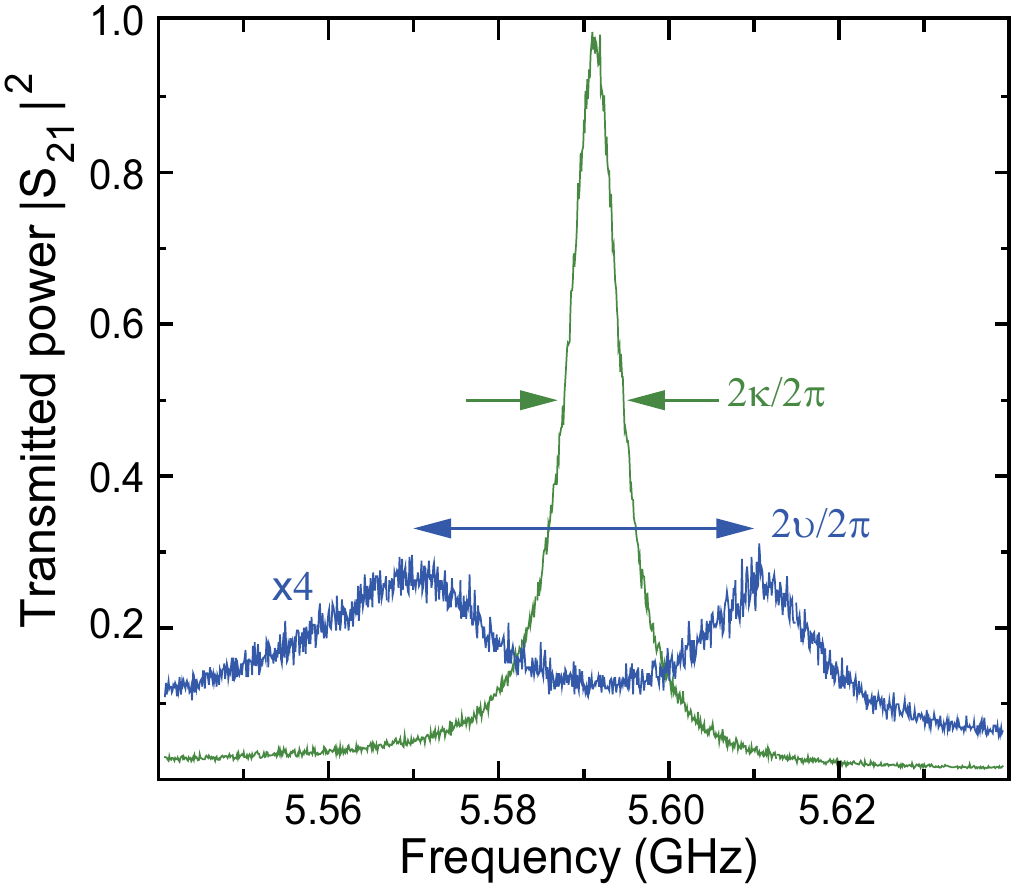}
    	\caption{(Color online)~Normal mode splitting for the spin transition $S_{2b}$ at $B=235$~mT extracted from the spectrum, which is presented in Fig.~\ref{fig_spectrum_main}, and the spectrum of the uncoupled resonator.}
	\label{fig_spectrum_ALC}
\end{figure}

A detailed study of the avoided level crossing at 235 mT is presented in Fig.~\ref{fig_spectrum_ALC}. The transmission spectra at the spin-cavity resonance shows a clear normal mode splitting. The fit of the experimental curve to the theory~\cite{Shuster2010} yields a coupling strength of $v/2\pi=21.2\pm0.3$~MHz and an inhomogeneous HWHM spin linewidth of $\Gamma^{\star}_2/2\pi=18\pm0.7$~MHz. An independent measurement of a similar 0.005\% Er:YSO crystal at a Bruker Elexsys 580 ESR spectrometer shows an inhomogeneous linewidth of approximately 5 MHz. Such a difference may be explained by two effects: the inhomogeneity of the magnetic DC field $B_0$ and mechanical stress exerted by the supporting teflon corks.

The HF spectrum of $^{167}$Er at the frequency region around 3-6 GHz is rather complex, which prevents us from the identification the magnetic transitions with sufficient confidence, see also Ref.~\cite{Bushev2011, Probst2013}. The typical coupling strength of the HF transitions around the $S_{2b}$ electronic spin transition is $v_{HF}/2\pi\simeq6$~MHz, with an inhomogeneous spin linewidth of $\Gamma^{\star}_{HF}/2\pi\simeq10$~MHz.

The presented values of the coupling strengths were obtained for an optimal orientation of the Er:YSO crystal with respect to the DC field. In general, the magnetic properties of the spins in the Er:YSO crystal is described by the \textbf{g}-tensor, which implies the dependence of g-factor and hence the magnitude of the Zeeman splitting on the relative orientation between the applied magnetic field and the crystal axes~\cite{Guillot2006}. The spin Hamiltonian of the erbium isotopes without a nuclear spin can be written as
\begin{equation}\label{SpinH}
H = \mu_B (\vec{B_0}+\vec{B_1} \cos \omega t) \cdot \textbf{g} \cdot \vec{S},
\end{equation}
where $\textbf{g}$ is the g-factor tensor, $\vec{B_0}$ is the applied DC magnetic field, $\vec{B_1} \cos \omega t$ is the oscillating (AC) field at angular frequency of $\omega$. The first term in the spin Hamiltonian in Eq.~(\ref{SpinH}) describes the Zeeman splitting, and the second term shows the anisotropy of the coupling strength, see also~\cite{Bertaina2009, Probst2013}. More explicitly, the coupling strength per single spin is determined by the resonant vacuum field strength $v_1 = \textrm{g}_{ac} \mu_B B_{\text{vac}}/\hbar$, where $g_{ac}$ is the effective g-factor of spins along the direction of oscillating vacuum magnetic field inside the cavity $\vec{B}_{\text{vac}} \cos \omega_c t$.

\begin{figure}[!h]
    	\includegraphics[width=0.9\columnwidth]{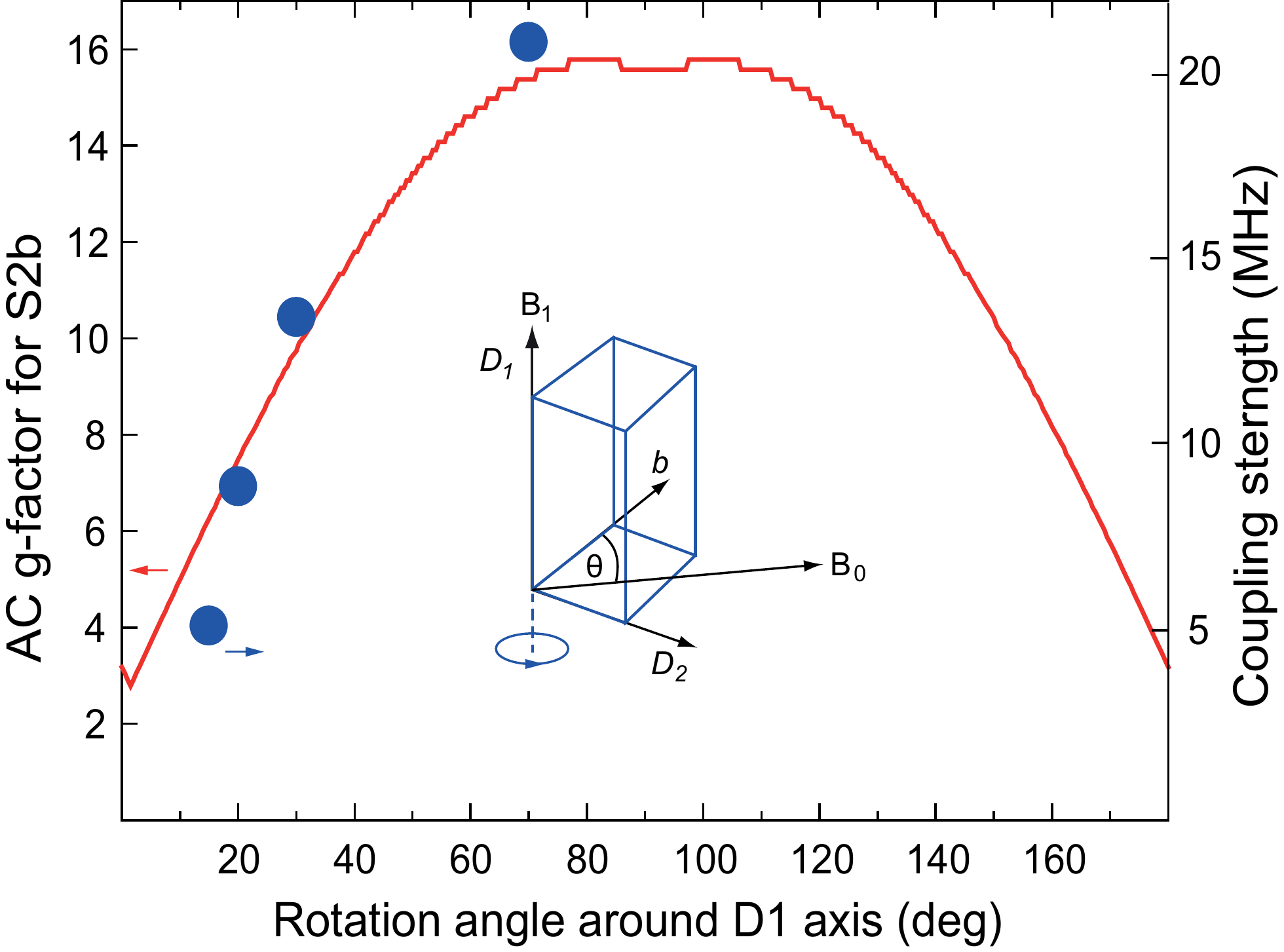}
    	\caption{(Color online)~Dependence of the coupling strength (circles) of the $S_{2b}$ transitions on the rotation angle $\theta$ around $D_1$ axis. The solid line is the simulated AC g-factor for the same transition.}
	\label{fig_angle}
\end{figure}

To explore the effect of the magnetic anisotropy of the Er:YSO crystal, we rotate the crystal around its $D_1$ axis and measure the dependence of the collective coupling strength $v/2\pi$ on the angle $\theta$ between $b$ and $D_2$ axis \cite{Sun2008}. The measured data is presented with filled circles, see Fig.~\ref{fig_angle}. We calculate $\textrm{g}_{ac}$ as a function of the angle $\theta$ using the EASYSPIN program~\cite{EasySpin}. Since the AC g-factor is proportional to the coupling strength, we superimpose it with the measured data. Due to the limited number of experimental points and imperfections in the angular alignment of the crystal, we can make only a qualitative comparison. The coupling strength increases from 4 MHz at 0$^{\circ}$ to 21 MHz at 70$^{\circ}$, which is accompanied by a 5-fold increase of $\textrm{g}_{ac}$.

For the further characterization of the sapphire loaded cavity, we estimate the coupling strength per single spin. Here, we neglect the spatial variations of the vacuum field of the TE$_{001}$ mode, since the dimensions of the crystal is much smaller than dimensions of the cavity. The number of spins participating in the magnetic microwave interaction is $N_s=n_s V\sim 8\cdot10^{15}$, where $V=0.05$~cm$^{3}$ is the volume of the crystal. The coupling strength per a single spin with $\textrm{g}_{ac}\approx15$ is $v_1=v/\sqrt{N_s}\approx 2\pi \times 0.24$~Hz. For a single spin with $\textrm{g}=2$, the corresponding coupling strength of the presented resonator is 0.03 Hz.

Having demonstrated coherent coupling to a solid state spin system, we would like to draw the reader's attention to the prospects of this open architecture. From the engineering point of view, 3D microwave resonators simplify the implementation of a microwave quantum memory based on trapped and laser cooled atoms. The 3D architecture can be a promising alternative for the implementation of coherent strong coupling of atomic clouds compared to planar circuits, as it has been proposed so far~\cite{Verdu2009}. Trapped atoms possess quite small inhomogeneous broadening which benefits the proposed setup. The electrical current distribution inside the resonator operating at the TE$_{001}$ mode allows for an easy transfer of atomic clouds into the cavity through holes. The setup will be similar to the frequency standards based on Cs fountains~\cite{RiehleBook}. For example, if $N_s\sim10^{10}$ Cs atoms are loaded inside the resonator with $\omega_c/2\pi \simeq 9.2$~GHz, then the collective coupling strength is estimated to be $v_1\sqrt{N_s}/2\pi \sim 5$~kHz. To reach the strong coupling regime in such an experiment one requires a high-Q resonator with $Q_l > 10^6$, which is certainly feasible by using superconducting cavities.

In conclusion, we have presented a cavity QED experiment with an Er$^{3+}$:Y$_2$SiO$_5$ crystal magnetically coupled to a 3D sapphire loaded cylindrical resonator. The magnetic coupling and the anisotropy was studied for different orientations of the crystal with respect to the DC and AC magnetic fields. Our experiments demonstrate that the strong coupling regime is attained for spin doped solids magnetically coupled to 3D resonators. We find a maximum collective coupling strength of 21.2 MHz with an inhomogeneous linewidth of 18 MHz. This work opens up at least two perspectives. First, quantum memories based on such solid state spin systems can be integrated with the emerging technology of 3D superconducting qubits. Second, the 3D architecture presents a promising alternative to a planar hybrid circuits coherently coupled to atomic clouds.

We thank S.~Poletto, O.~Mishina, J.~Eschner and S.~Ritter for the stimulating discussion and critical reading of the manuscript. S.~P. acknowledges financial support by the LGF of Baden-W\"{u}rttemberg. This work was supported by Australian Research Council Grant No. CE11E0082 and FL0992016 and the BMBF program "Quantum communications".

\bibliographystyle{apsrev}

\bibliography{Er_3D}

\end{document}